\documentclass[aps,pra,reprint,floatfix,showpacs,nofootinbib,longbibliography]{revtex4-1}
\usepackage{amssymb,amsmath,graphicx,color}
\usepackage[colorlinks=true,urlcolor=blue,citecolor=blue]{hyperref}

\newcommand{\ignore}[1]{}

\begin{document}

\title{High-Resolution Spectroscopy of {R}ydberg Molecular States of $^{85}${R}b$_2$ Near the $5s+7p$ Asymptote}

\author{R.A. Carollo}\altaffiliation{Present address: Department of Physics, Amherst College, Amherst, MA 01002-5000, USA}
\author{J.L. Carini}\altaffiliation{Present address: 400 Main St., East Hartford, CT 06118, USA}
\author{E.E. Eyler}\thanks{Deceased September 19, 2016}
\author{P.L. Gould}
\author{W.C. Stwalley}
\affiliation{Department of Physics, University of Connecticut, Storrs, CT 06269}

\date{\today}


\begin{abstract}
Rydberg molecules, often exemplified by long-range ``trilobite'' molecules, are a subject of much recent interest at high principal quantum number $n$. States that use the same bonding mechanism can exist at much lower $n$ and less-extreme internuclear separations that are still quite long range. We use a high-resolution pulsed amplifier to study previously-detected transitions to a low-$n$ Rydberg molecular state near the $5s+7p$ asymptote in $^{85}$Rb$_2$. The observed line structure is modeled using precise ground-state positions and a lower bound of $\tau > 1.1 \times 10^{-9}$ s is set on the excited-state autoionization lifetime.
\end{abstract}

\maketitle

\section{Introduction}

Long-range Rydberg molecules are a class of molecules formed by an unusual bonding mechanism. This mechanism was introduced by Du and Greene in Ref.~\cite{greene87} and extended by Greene \emph{et al.} in Ref.~\cite{greene00}. It results in two types of long-range Rydberg states, including ``trilobite'' molecules (perturbed hydrogenic states), nicknamed for the resemblance of the electron wavefunction to trilobite fossils. These molecules are characterized by the $s$-wave scattering of a Rydberg electron on a nearby ground-state atom. There are two classes of these molecules: low-$\ell$ ($\ell \leq 2$) and high-$\ell$. The high-$\ell$ molecules are the true trilobites, with deep potentials and a large permanent dipole moment on the order of a kiloDebye. They are actually composed of multiple quasidegenerate high-$\ell$ states. There is a similar class of molecules, also with deep potentials and a large permanent dipole moment, termed ``butterfly'' molecules, that result from $p$-wave scattering of Rydberg electrons~\cite{sadeghpour02}.

The low-$\ell$ states of both $s$-wave and $p$-wave scattering types have shallower potentials with a smoother shape. The interaction is proportional to the square of the electron wavefunction for $s$-wave scattering, and to the square of the gradient of the electron wavefunction for $p$-wave scattering~\cite{rost10}:

\begin{equation}
V(R) = a_s(R) |\psi_e(R)|^2 + 6 \pi a_p^3(R) |\nabla \psi_e(R)|^2 ,
\end{equation}
where $a_s$ and $a_p$ are the $s$-wave and $p$-wave electron scattering lengths, respectively.

Several other groups are doing experiments with long-range Rydberg molecules, including true trilobite molecules. Of particular note are the groups of Shaffer and Pfau. Confirmation of the existence of these states at low $n$ was derived from data taken in a high-temperature Rb heat-pipe oven in 2006~\cite{niemax06}. The creation of higher-$n$ long-range Rydberg molecules at ultracold temperatures was achieved by a 2009 collaboration of the Shaffer and Pfau groups, also in Rb~\cite{pfau09}. A permanent dipole moment in these states was measured in 2011~\cite{pfau11}. Larger dipole moments (up to 100 Debye) were measured in Cs$_2$ in 2012~\cite{shaffer12}. True high-$\ell$ trilobite molecules were formed in Cs by Booth \emph{et al.} in 2015~\cite{shaffer15}. Most recently, Niederpr{\"u}m \emph{et al.} detected high-$n$ butterfly states in $^{87}$Rb~\cite{ott16}. Other important related work includes Refs.~\cite{raithel14,burgdorfer15,greene16}. Theoretical work has also been continued on these exotic molecular states~\cite{greene15}.

Our own ultracold observations~\cite{bellos13,bellos13a} of low-$n$ (7, 9-12) and low-$\ell$ long-range $^{85}$Rb$_2$ Rydberg molecules involved similar states to those studied in the high temperature work of~\cite{niemax06}. This is an interesting regime since the bonding is not due solely to low-energy electron scattering, but also has some covalent character. We scanned a region of over 100 cm$^{-1}$ with a frequency-doubled pulsed dye laser at a resolution of $\sim 1$ cm$^{-1}$ to obtain a spectrum.

In addition to the lower $n$ that we study compared to the Pfau and Shaffer groups, we also use a technique with other important differences. The work of Bendkowsky \emph{et al.} involves two-photon photoassociation of $^{87}$Rb that is prepared in an Ioffe-Pritchard trap at a density of $1.5 \times 10^{13}$ cm$^{-3}$ and a temperature of $3.5 \, \mu$K~\cite{pfau09}. The experiment of Booth \emph{et al.} utilizes a far off-resonance trap (FORT) of Cs with a density of $5 \times 10^{13}$ cm$^{-3}$ and a temperature of $40 \, \mu$K~\cite{shaffer15}. The densities in both of these experiments are more than two orders of magnitude higher than our magneto-optical trap (MOT) ($1 \times 10^{11}$ cm$^{-3}$) and, in the Rb apparatus of Bendkowsky \emph{et al.},  1.5 orders of magnitude colder (our MOT is $100 \, \mu$K). By exciting ultracold bound molecules produced by photoassociation to long-range Rydberg states, rather than starting with free atoms, we are able to produce them at densities and temperatures that are technically simpler to achieve.

In this paper (derived from Ch. 5 of Ref.~\cite{carollo15}), we re-examine selected lines from our previous study at higher resolution. We utilize a pulsed amplifier in order to achieve a narrow linewidth to identify structure and set a limit on the autoionization rate of the long-range Rydberg molecules.

\section{Experiment}

The general design of our experiment is similar to our earlier work~\cite{bellos13,bellos13a}. Specific details relevant to the present work will be expanded below.

\begin{figure}[tbh]
\begin{center}
\includegraphics[width=\columnwidth]{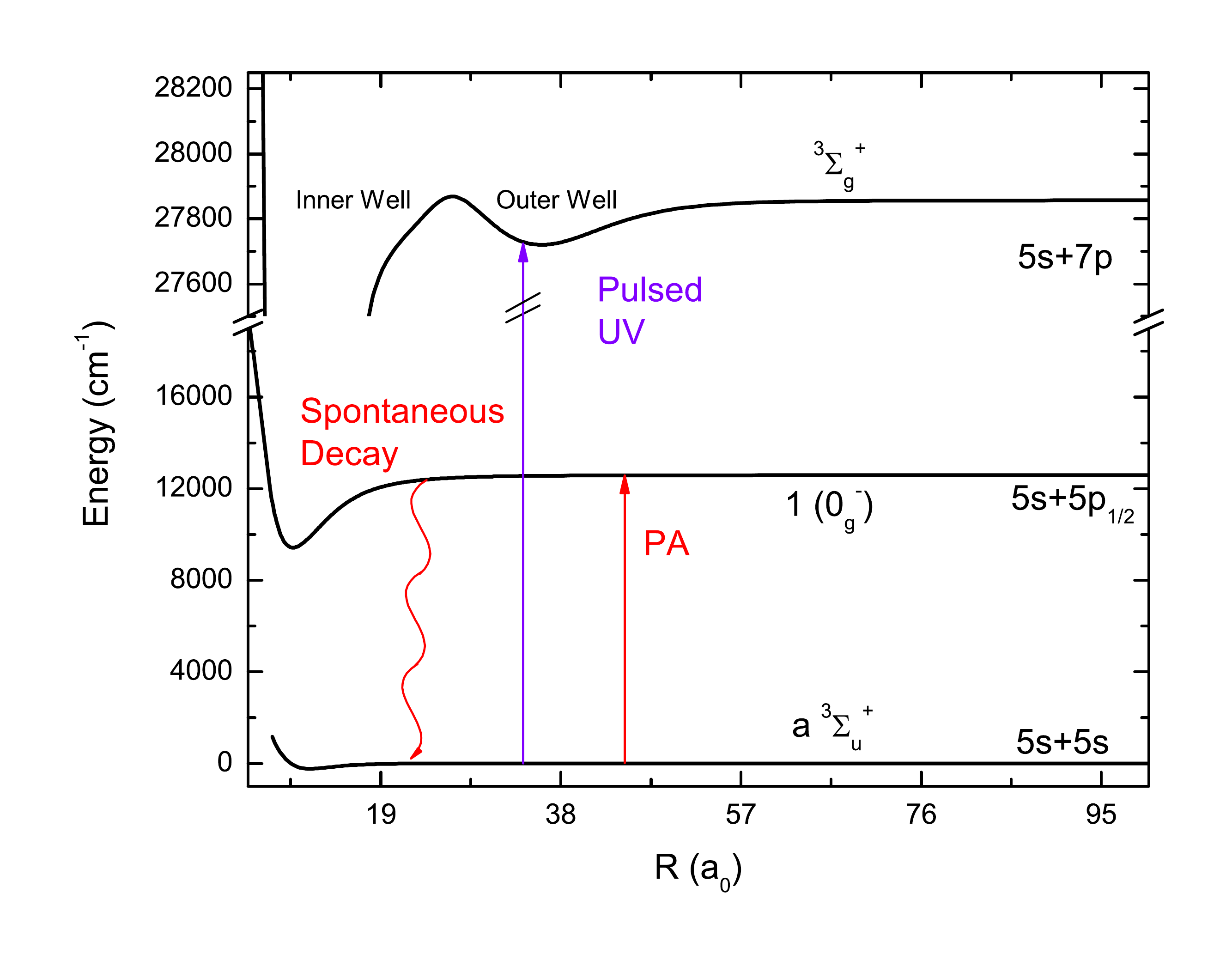}
\caption[The experimental scheme for long-range Rydberg molecule production]{[Color online] The experimental scheme for long-range Rydberg molecule production. We photoassociate to $v' \simeq 173$ of the $1 \, (0_g^-)$ state below the $5p_{1/2}$ asymptote. This level spontaneously decays to the upper vibrational levels of the $a \, ^3 \Sigma_u^+$ state, especially $v'' = 35$ and 36. After these precursor  molecules are formed, they are excited to the long-range Rydberg $^3 \Sigma_g^+$ state near the atomic $7p$ asymptote.}
\label{rydberg:formation}
\end{center}
\end{figure}

A schematic representation of our experimental procedure is shown in Fig.~\ref{rydberg:formation}. After cooling atoms in a MOT to $\sim 100 \, \mu$K we perform photoassociation to $v' \simeq 173$ of the $1 \, (0_g^-)$ state~\cite{huang06a} at $12,561.8$ cm$^{-1}$\ignore{(see Fig.~\ref{trap-loss})}. As we have full rotational resolution, we typically photoassociate to the $J' = 1$ line. The photoassociation laser is a fiber-coupled continuous-wave (CW) Coherent 899-29 Ti:Sapphire laser (medium-wave optics) that delivers 450 mW of optical power to the vacuum chamber. This state decays primarily to the $v'' = 35$ and 36 levels of the $a \, ^3 \Sigma_u^+$ state. These precursor molecules are then excited to the target $^3 \Sigma_g^+$ state levels near the $5s + 7p$ asymptote via a frequency-doubled nanosecond pulsed amplifier.

The pulsed amplifier was originally designed for earlier Rydberg work~\cite{eyler94}. It is composed of three Bethune dye cells~\cite{bethune81} that amplify a fiber-coupled Ti:Sapphire laser (899-29, modified short-wave optics). The Ti:Sapphire beam sets the frequency that the dye cells amplify, acting in place of a dye-cell oscillator such as our ND6000 dye laser uses. The Ti:Sapphire CW power must be sufficiently high (above $\sim 100$ mW) at the first cell to saturate the amplifier. These cells are pumped by a Spectra-Physics Quanta-Ray (LAB-150) injection-seeded Nd:YAG laser. The Nd:YAG injection seed is an OEM option that both stabilizes the frequency and the power envelope of each pulse. Spectra-Physics specifies a reduction in the second-harmonic 532 nm linewidth from $< 1.0$ cm$^{-1}$ unseeded to  $< 0.003$ cm$^{-1}$ with the injection seeder~\cite{lab150}. Injection seeding the Nd:YAG is vital to achieving stable, narrow pulse bandwidth. The pulse-amplified bandwidth of second-harmonic radiation at $\sim 360$ nm generated by a potassium dihydrogen phosphate (KDP) crystal was measured by scanning through the atomic $7p_{1/2}$ line at low pulse energies. This resulted in a measured linewidth of $\lesssim 150$ MHz\ignore{(see Fig.~\ref{linewidth})}. Typical UV pulse energies are 0.4 mJ per pulse\ignore{(see Fig.~\ref{energyhistogram})}.

\begin{figure}[tbh]
\begin{center}
\includegraphics[width=\columnwidth]{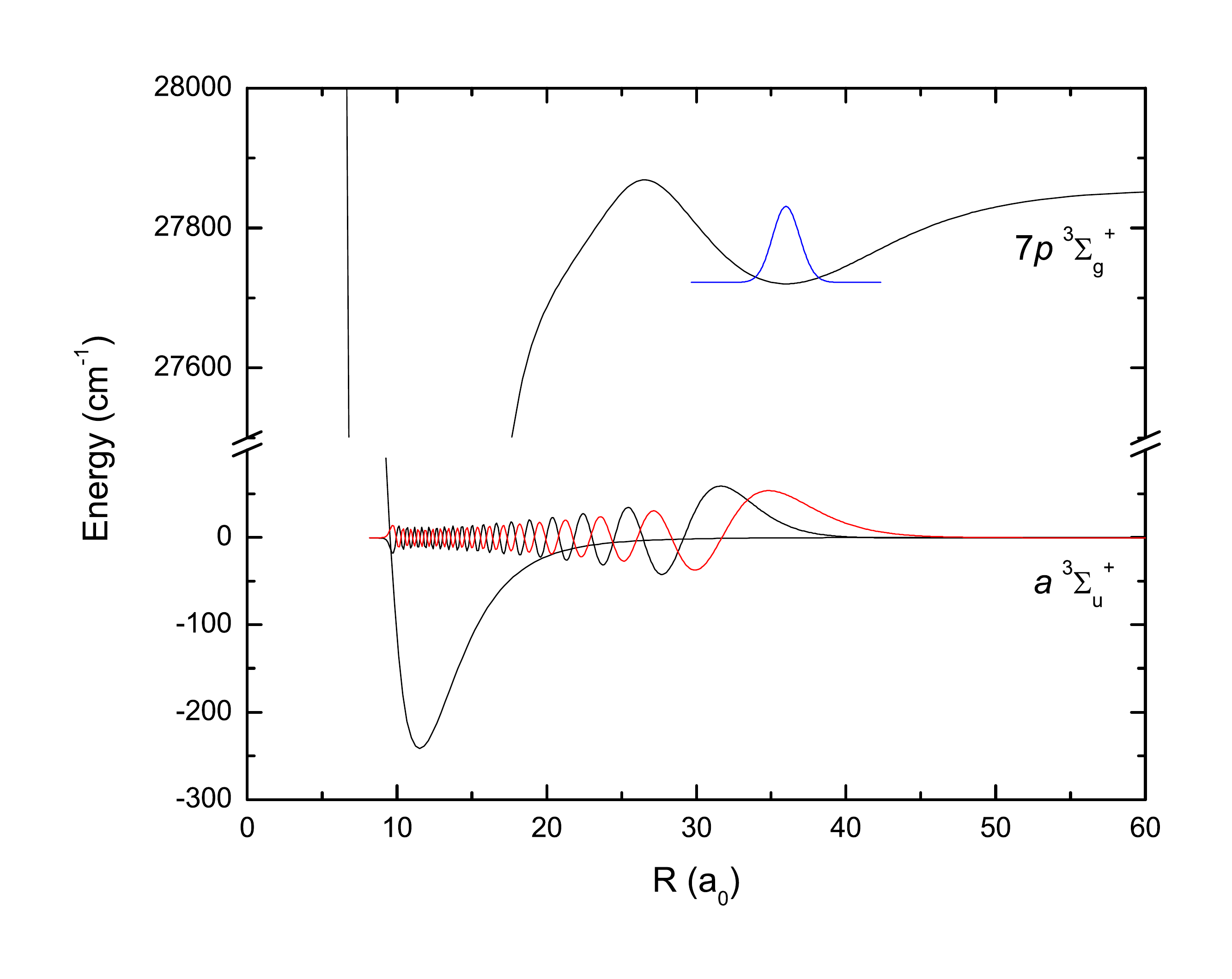}
\caption[Wavefunctions of the  $a \, ^3 \Sigma_u^+$ state $v'' = 35$ and 36 and Rydberg $7p$ $^3 \Sigma_g^+$ state outer well $v' = 0$]{[Color online] Wavefunctions of the  $a \, ^3 \Sigma_u^+$ state $v'' = 35$ (black) and $v'' = 36$ (red) and Rydberg $7p$ $^3 \Sigma_g^+$ state outer well $v' = 0$.}
\label{rydberg:wavefnct}
\end{center}
\end{figure}

The high-lying vibrational levels we create in the $a \, ^3 \Sigma_u^+$ state are ideal for accessing the long-range Rydberg state that correlates to the atomic $7p$ asymptote. The outer turning points of these wavefunctions match the position of the $^3 \Sigma_g^+$ outer well $v' = 0$ state very well, as shown in Fig.~\ref{rydberg:wavefnct}. The $v' = 35$ wavefunction, shown in black, has an outer turning point maximum amplitude at $\sim 30$ a$_0$, and the $v' = 36$ wavefunction, in red, has a maximum amplitude in the final lobe at $\sim 35$ a$_0$. The \emph{ab initio} potential shown in Fig.~\ref{rydberg:wavefnct} and used to generate the displayed wavefunctions is from Ref.~\cite{bellos13}.

After excitation, the long-range Rydberg molecules autoionize into a Rb$_2^+$ ion and a free electron. The Rb$_2^+$ ions are collected by our ion optics and, after a free-flight region, are detected by a discrete dynode multiplier (ETP model 14150). The detected signals are mass-selected for dimer ions by time-of-flight mass spectrometry and integrated by a boxcar integrator (SRS SR250).

To reduce contamination of the molecular channel by atomic ions, the MOT lasers are turned off for $10 \, \mu$s before and after each UV laser pulse. With the acousto-optical modulator (AOM) switched off, the trap laser extinction is 60 dB, with 30 mW of power at the fiber output.

All spectra shown in this chapter were obtained using several forms of averaging. The inner-well states were acquired using 10-shot exponential averaging on the boxcar integrator. The final spectra are composed of multiple scans that were averaged using the custom-written Mathematica package provided as Supplemental Material. The outer-well states were recorded using single-shot sampling from the boxcar integrator; the resulting spectra were then filtered using Mathematica's low-pass Fourier filter. This type of post-processing filter does not distort the line shape of spectral features. Multiple scans were again averaged as above, with the addition of supplementary linearization, also provided in the Supplemental Material.

The reported frequencies for the spectra are laser frequency, not the absolute energy of the spectral features. There may also be rare errors on the order of 0.22 cm$^{-1}$ due to the design of our 899-29 wavemeter.

\section{Low-Resolution Data and Franck-Condon Factors}

\begin{figure}[tbh]
\begin{center}
\includegraphics[width= \columnwidth]{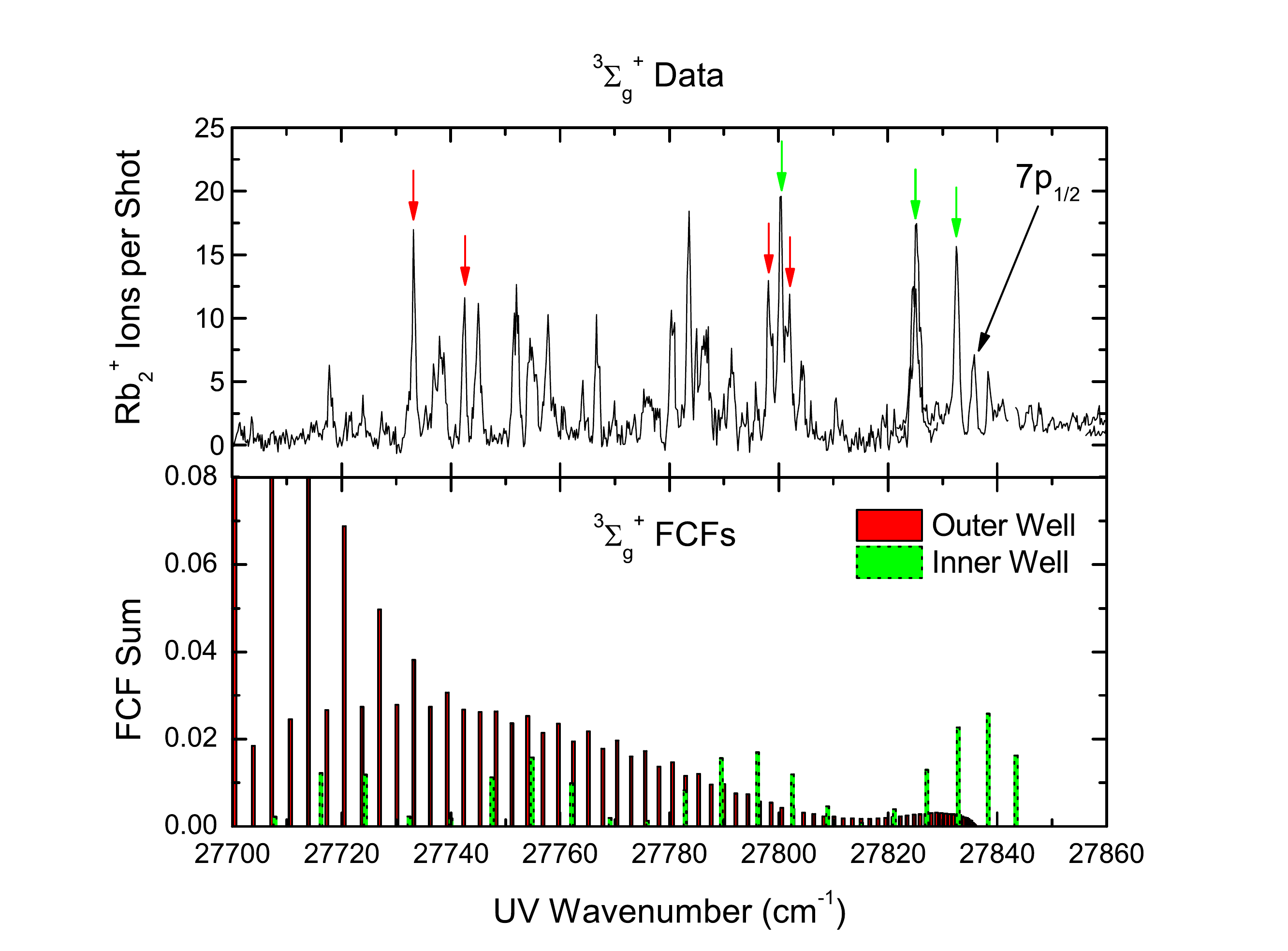}
\caption[An overview showing FCFs to the Rydberg potential from our precursor molecules and comparing them to low-resolution data]{[Color online] The lower panel of this figure shows the sum of FCFs for transitions to the long-range Rydberg $^3 \Sigma_g^+$ potential from the $a \, ^3 \Sigma_u^+$ $v'' = 35$ and 36 levels. Transitions to the inner well are shown in green, while those to the outer well are shown in red. The upper panel shows our low-resolution data from Ref.~\cite{bellos13a}. Spectral regions that were studied at higher resolution are marked with red arrows for outer-well levels, shown in Figs.~\ref{rydberg:outer1} and~\ref{rydberg:outer2}. The corresponding regions for inner-well levels are marked with green arrows, shown in Figs.~\ref{rydberg:hyperfine_comparison},~\ref{rydberg:jcomparison}, and~\ref{rydberg:long} (two weak outer-well levels are also shown in Fig.~\ref{rydberg:long}).}
\label{rydberg:overview}
\end{center}
\end{figure}

As seen in Fig.~\ref{rydberg:wavefnct}, the overlap for excitation to the bottom of the outer well of the long-range Rydberg state is quite good. As expected, this translates into large Franck-Condon factors (FCFs). The factor, FCF $ = \left| \left<\psi_{v'} | \psi_{v''} \right> \right|^2$, is typically a good approximation for the transition probability. Despite the good approximation the FCF provides, adding the transition dipole moment to the calculation would be beneficial, but unfortunately the transition dipole moment from the $a \, ^3 \Sigma_u^+$ potential to the $7p \, ^3 \Sigma_g^+$ potential is currently unknown. The FCFs for transitions from $v'' = 35$ and $v'' = 36$ with $J'' = 0$ to the various levels of the inner and outer wells were calculated using LEVEL 8.2~\cite{leroylevel16} and the potential shown in Fig.~\ref{rydberg:wavefnct} and can be found in two tables included in the Supplemental Material. The sum of these FCFs is shown in Fig.~\ref{rydberg:overview}, on the same horizontal axis as our low-resolution data from~\cite{bellos13a}.

\begin{figure}[tbh]
\begin{center}
\includegraphics[width=\columnwidth]{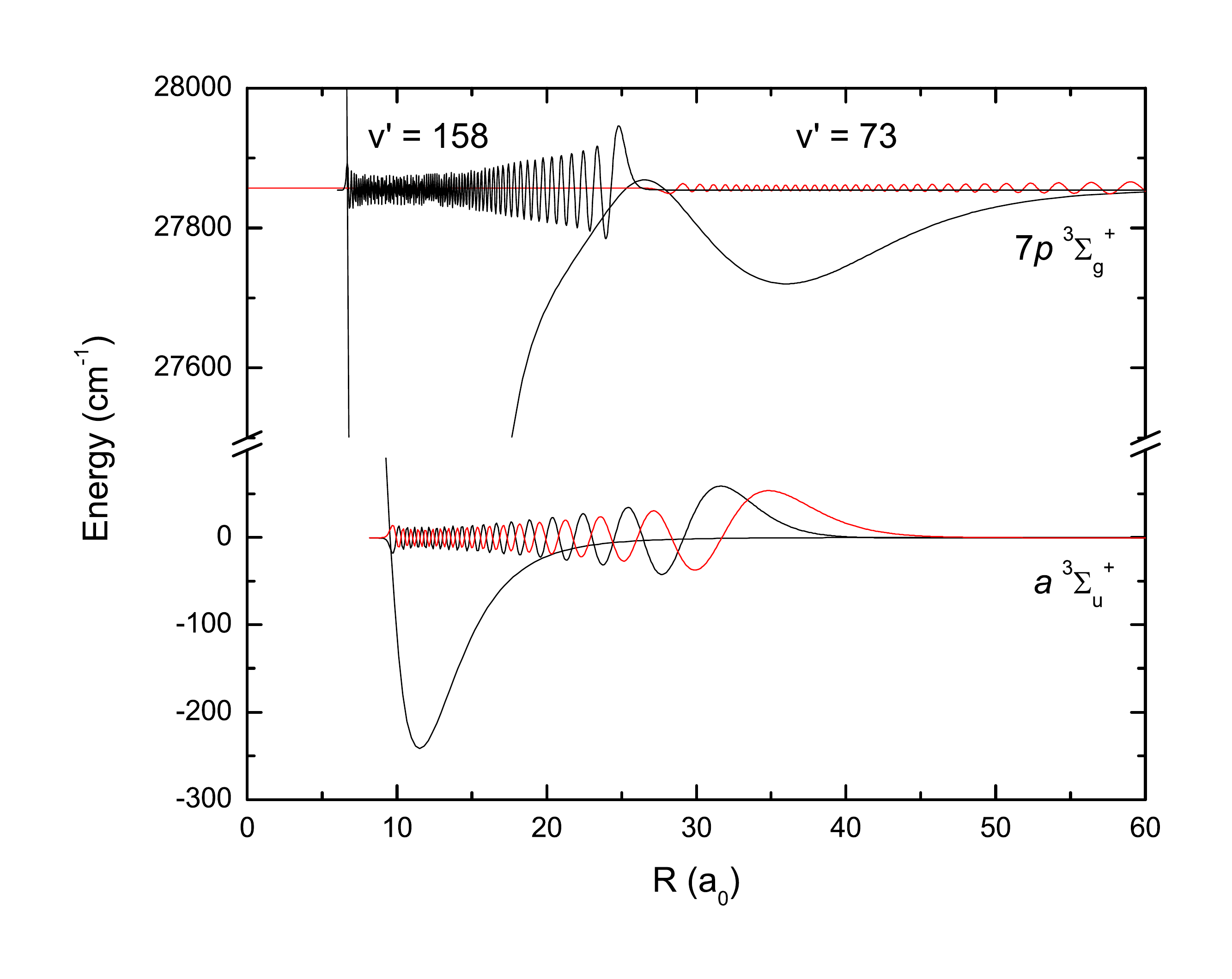}
\caption[Wavefunctions of the  $a \, ^3 \Sigma_u^+$ state $v'' = 35$ and 36 and $7p$ $^3 \Sigma_g^+$ state outer well $v' = 73$ and inner well $v' = 158$]{[Color online] Wavefunctions of the  $a \, ^3 \Sigma_u^+$ state $v'' = 35$ (black) and $v'' = 36$ (red) and Rydberg $7p$ $^3 \Sigma_g^+$ state outer well $v' = 73$ and inner well $v' = 158$.}
\label{rydberg:wavefnct2}
\end{center}
\end{figure}

It should be noted that, although we discuss the inner well and outer well of the $^3 \Sigma_g^+$ state, they are actually a single system. What we have designated as an inner well or outer well level is nonetheless an eigenstate of the entire potential. Tunneling through the barrier, although small, does allow for some wavefunction amplitude to leak into the other well, especially near dissociation. The highest bound state of the system, $v' = 232$ (or $v' = 73$ of the outer well), has some amplitude in the inner well, but its amplitude in the outer well is $\sim 422$ times greater. The highest bound level of the inner well, $v' = 219$ ($v' = 158$ of the inner well) has $1,890$ times more amplitude in the inner well than in the outer well. The amplitude is measured as the peak to trough height of the first oscillation on either side of the potential barrier. These vibrational levels are shown in Fig.~\ref{rydberg:wavefnct2}.

Fig.~\ref{rydberg:overview} also has several arrows indicating particular vibrational levels. At higher energies, $27,800$ cm$^{-1}$ and above, the three green arrows indicate lines we have measured at high resolution and have assigned to the inner well. At lower energies, marked in red near $27,730$ cm$^{-1}$, are two more regions in which we have observed high-resolution spectra assigned to the outer well. Two additional outer-well states are observed at $\sim 27,800$ cm$^{-1}$, but are weak and poorly-resolved.

The $^3 \Sigma_g^+$ potential that we used to calculate these FCFs, from Ref.~\cite{bellos13}, is not calculated in a spin-orbit coupled basis, and thus has inherent inaccuracies when compared to empirical data. The spin-orbit splitting between the $7p_{1/2}$ and $7p_{3/2}$ in $^{85}$Rb is 35 cm$^{-1}$, so we should expect uncertainties of that scale. The $a \, ^3 \Sigma_u^+$ potential used for these calculations is the semi-empirical potential of Ref.~\cite{tiemann10} and is considered to be highly accurate.

Although the spin-orbit splitting was not taken into account in the calculation, we can still identify to which asymptote certain states correspond. In Fig. 5 and Tab. 1 of Ref.~\cite{stwalley12}, we see that, at the $5p$ asymptotes, all of the components of the $^3 \Sigma_g^+$ state correspond to the $5p_{1/2}$ limit, and all components of the $^3 \Pi_g$ state correspond to the $5p_{3/2}$ limit. Thus we are confident that the $^3 \Sigma_g^+$ and $^3 \Pi_g$ states of the $7p$ asymptote will similarly correlate to the $7p_{1/2}$ and $7p_{3/2}$ asymptotes, respectively.

There are several interesting features of the FCFs. The first is that the inner and outer wells each dominate in a different energy range, although the change in strength of the inner-well FCFs is less significant. The inner well has the highest excitation probability for the highest several vibrational levels of the Rydberg molecular state, near the atomic asymptote. The outer well has very low excitation probability near the atomic asymptote, but quite high excitation probability at the bottom of the well.

A second feature of note is the strong even-odd oscillation in outer-well FCFs at low energy. This effect is strongly dependent on the $R$ value of the outer turning point, and can be seen in the difference between the FCFs from $v'' = 35$ and $v'' = 36$ in Fig.~\ref{rydberg:fcf35} and Fig.~\ref{rydberg:fcf36}, respectively. Full tables of these FCFs are in the Supplemental Material.

\begin{figure}[tbh]
\begin{center}
\includegraphics[width=\columnwidth]{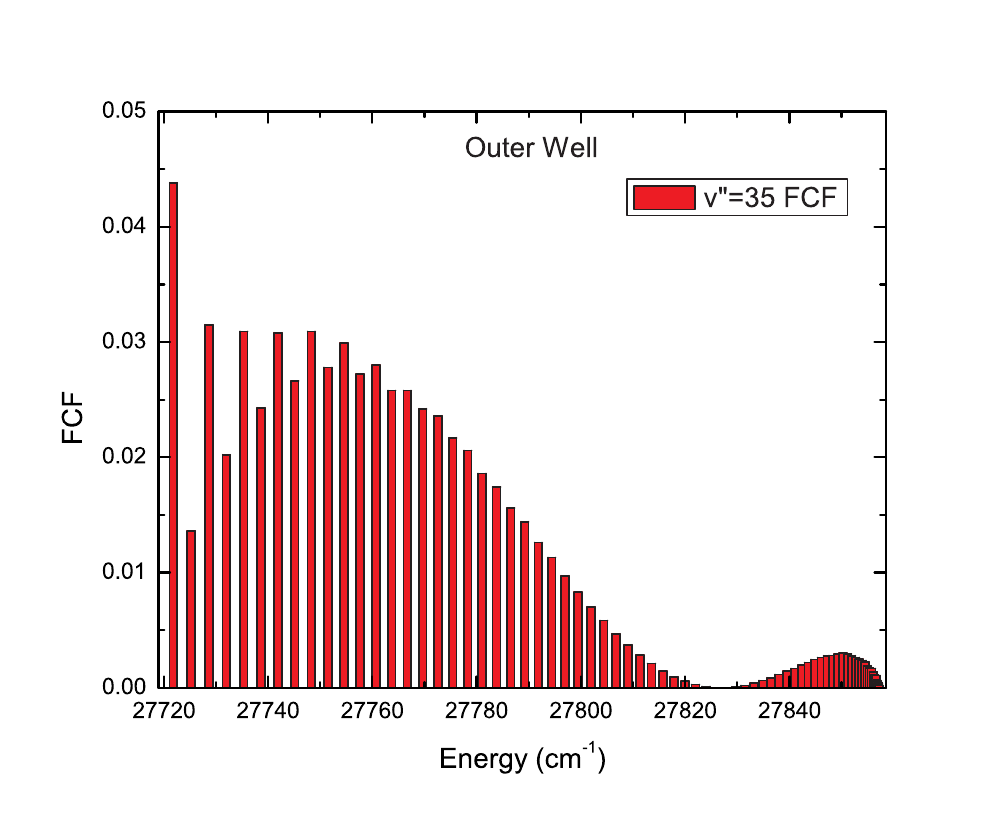}
\caption[FCFs to the outer well of the Rydberg molecular potential, from $a \, ^3 \Sigma_u^+$ $v'' = 35$.]{[Color online] The Franck-Condon Factors for transitions from the $a \, ^3 \Sigma_u^+$ $v'' = 35$ to the Rydberg molecular potential are shown. The maximum FCF is $4.4 \%$ for $v' = 0$.}
\label{rydberg:fcf35}
\end{center}
\end{figure}

Fig.~\ref{rydberg:fcf35} shows some oscillations in the $v'' = 35$ FCFs at low $v'$, and also shows a maximum FCF of $4.4 \%$ that is reached at $v' = 0$. Recalling the good wavefunction overlap with $v' = 0$ shown in Fig.~\ref{rydberg:wavefnct}, this is unsurprising.

In Fig.~\ref{rydberg:fcf36}, we see much stronger even-odd oscillations at the bottom of the well, and weaker oscillations extending toward the atomic asymptote. We also see a dramatically larger FCF in the lowest several even levels, of up to $38.7 \%$ for $v' = 0$, which is the source of the vast majority of the FCF strength seen in Fig.~\ref{rydberg:overview}. This is due to the extremely favorable overlap of the outer turning point of $v'' = 36$ with $v' = 0$ of the outer well.

\begin{figure}[tbh]
\begin{center}
\includegraphics[width= \columnwidth]{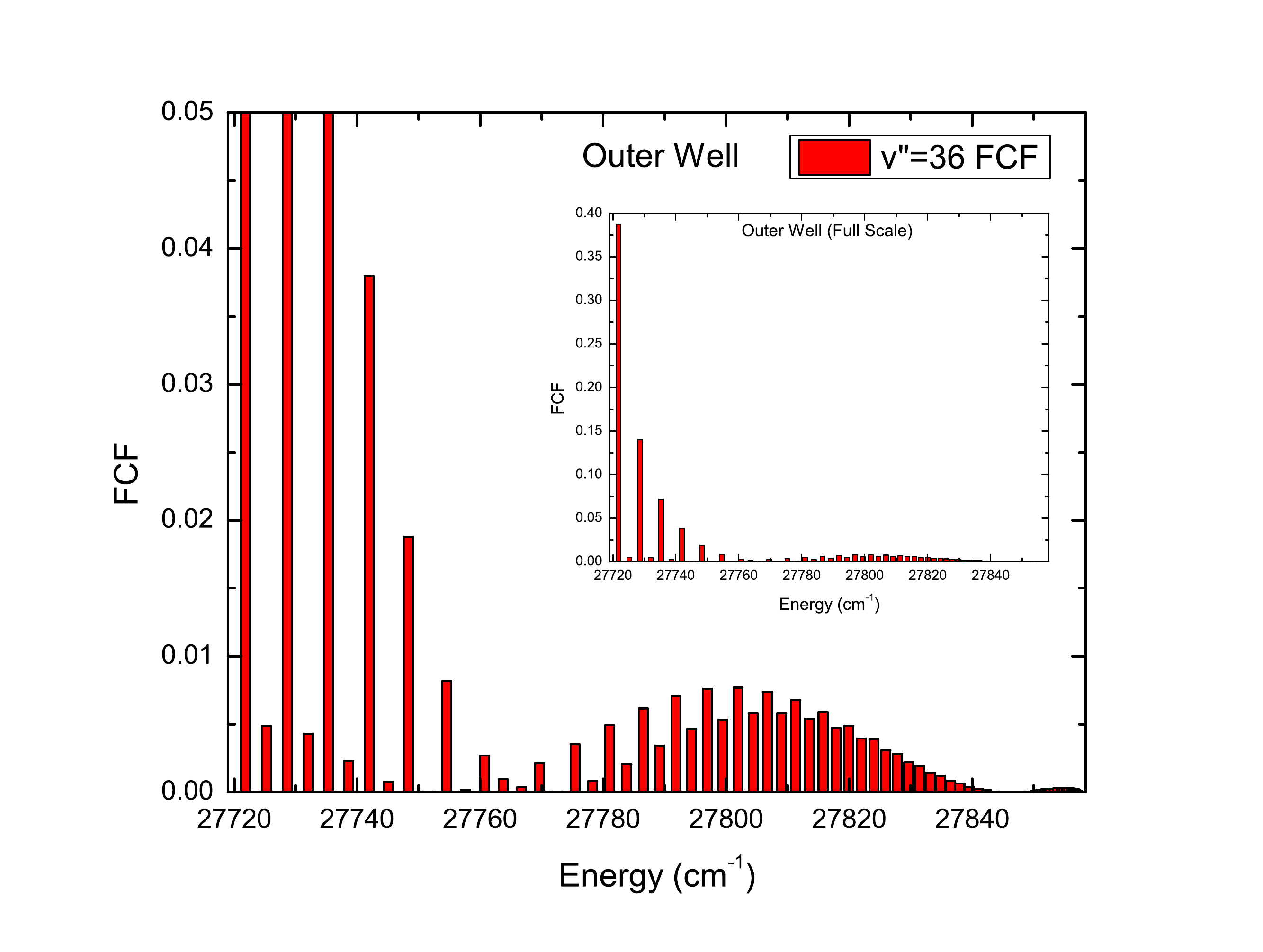}
\caption[FCFs to the outer well of the Rydberg molecular potential, from $a \, ^3 \Sigma_u^+$ $v'' = 36$.]{[Color online] The Franck-Condon Factors for transitions from the $a \, ^3 \Sigma_u^+$ $v'' = 36$ to the Rydberg molecular potential are shown. The scale of the main figure is the same as in Fig.~\ref{rydberg:fcf35}, but the inset shows the full scale of the lower FCFs. Note the dramatic increase in probability at low even values of $v'$ (to $38.7 \%$ for $v' = 0$) when compared with the FCFs in Fig.~\ref{rydberg:fcf35}}
\label{rydberg:fcf36}
\end{center}
\end{figure}

The cause of the even-odd oscillation in the FCFs is explained in detail in Ref.~\cite{carollo16}. It is important to note that the outer well is, like many potentials, fairly harmonic near the bottom. Thus its deeply-bound eigenstates are also fairly similar to the eigenstates of a harmonic oscillator. In an ideal harmonic oscillator, the $v' = 1$ wavefunction has one node, which will be located at the same $R$ as the maximum in the wavefunction of the $v' = 0$. The location of this node means that a fairly-localized wavefunction (in a lower electronic potential) that has good overlap with the $v' = 0$ state must have poor overlap with $v' = 1$, and thus a small FCF.

Similarly, the $v' = 2$ wavefunction, which has two nodes, will have an extremum located at the same $R$ as the maximum of $v' = 0$. (extremum rather than maximum because the sign of the wavefunction may flip depending on the phase convention. Nonetheless, $|\psi|^2$ will have a maximum.) This lobe of the wavefunction will have a lower amplitude than in $v' = 0$, and the two flanking lobes will partially cancel the overlap, resulting in a weaker but still-strong FCF. This process continues up the roughly harmonic well, with the alternation in strength gradually losing contrast. Anharmonicity in the real potential, other lobes in the lower wavefunction, and imperfect alignment of the outer lobe all further reduce the alternation. Because the outermost maximum of the $a \, ^3\Sigma_u^+$, $v'' = 36$ level is near $R_e'$ of the $^3 \Sigma_g^+$ outer well potential, but the maximum of the $v'' = 35$ level is well inside $R_e'$ of the $^3 \Sigma_g^+$ outer well, the contrast ratio between even and odd FCFs is much greater ($79.8 \times$ \emph{vs.} $3.2 \times$ for the $v' = 0$:1 ratio) in the $v'' = 36$ plot (Fig.~\ref{rydberg:fcf36}) than in the $v'' = 35$ plot (Fig.~\ref{rydberg:fcf35}).

\section{High-Resolution Scans}

In our previous work on these states~\cite{bellos13a}, we relied on REMPI measurements from the thesis of Ye Huang~\cite{huang06} to identify the distribution of $a \, ^3 \Sigma_u^+$ state molecules, and believed that most of the molecules were in $v'' = 35$. With our pulse-amplified CW laser providing a linewidth of $\sim 150$ MHz, or $\sim 200 \times$ narrower than our previous work, we can detect the ground-state distribution more accurately than it has been measured in our apparatus before.

\begin{figure}[tbh]
\begin{center}
\includegraphics[width= \columnwidth]{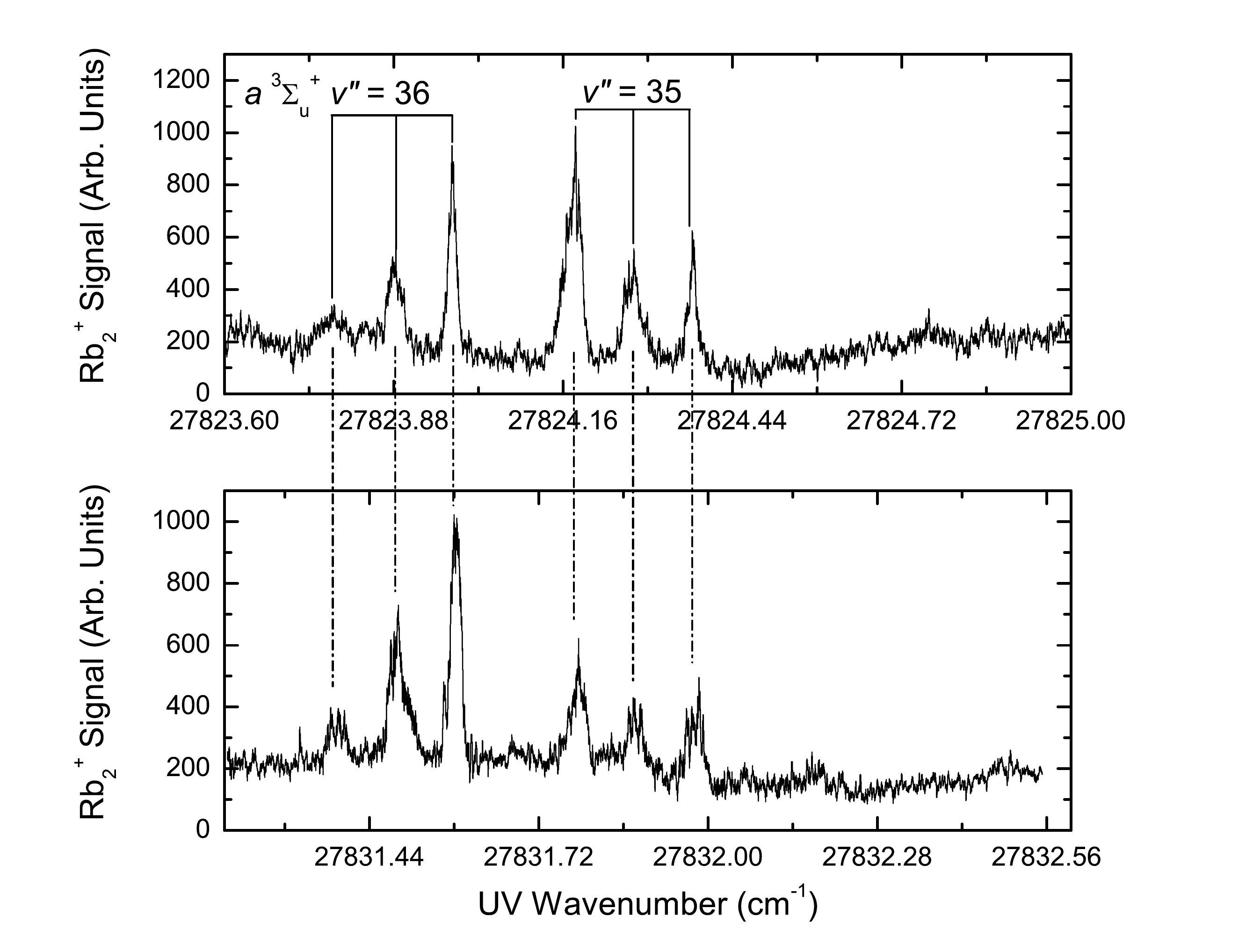}
\caption[Two inner-well spectra, aligned to show the identical $a \, ^3 \Sigma_u^+$ state hyperfine splitting]{Two inner-well spectra are shown, aligned to demonstrate their similarities. Both lines show the same vibrational and hyperfine structure from the $a \, ^3 \Sigma_u^+$ state. They also show roughly comparable relative intensities among the $v'' = 35$ and $v'' = 36$ lines.}
\label{rydberg:hyperfine_comparison}
\end{center}
\end{figure}

In Fig.~\ref{rydberg:hyperfine_comparison}, two sets of inner-well lines are shown (in Fig.~\ref{rydberg:overview}, these are indicated by the two highest-energy green arrows). Each spectrum contains two prominent triplets of lines. According to calculations from LEVEL, the $a \, ^3 \Sigma_u^+$ state $v' = 35$ level is bound by 0.8065 cm$^{-1}$, and the $v' = 36$ level is bound by 0.4161 cm$^{-1}$ (see Table B.2 in Appendix B of~\cite{carollo15} for a full listing of vibrational levels). The difference of $\sim 0.39$ cm$^{-1}$ matches the spacing from the first line of one triplet to the first line of the next, and so on for the second and third components of the triplets. Instead of our $a \, ^3 \Sigma_u^+$ state population residing primarily in the $v'' = 35$ level as we previously thought, we appear to have a nearly equal balance of $v'' = 35$ and $v'' = 36$ molecules. Even this observation, though, is affected by the FCFs of each $a$-state vibrational level, and the apparent line strength will shift with $v'$. For example, Fig.~\ref{rydberg:long} shows a very different ratio of $v'' = 35$ to 36 at an inner-well vibrational level that is $\sim 25$ cm$^{-1}$ more deeply bound than the level shown in the upper panel of Fig.~\ref{rydberg:hyperfine_comparison}.

Similarly, within each triplet the spacing between the first and second lines is the same as the spacing between the second and third. This spacing is consistent with the atomic $^{85}$Rb ground-state hyperfine splitting between $F = 2$ and $F = 3$ of $0.1012$ cm$^{-1}$~\cite{steck85}. The lowest-energy component (which must come from the highest-lying ground-state hyperfine level) must be the $\left| F = 3 \right> + \left| F = 3 \right>$ asymptote. The next highest corresponds to $\left| F = 2 \right> + \left| F = 3 \right>$, and the highest-energy component of the triplet corresponds to the $\left| F = 2\right> + \left| F = 2 \right>$ atomic asymptote.

Traditionally, it is assumed that photoassociation acts only on the electron orbital degrees of freedom, not altering electronic or nuclear spin~\cite{jones06}. Nuclear spin \emph{must} flip, however, to explain the states we see. Our MOT selects primarily for the $F = 3$ state, although a small $F = 2$ population can be observed in spectroscopy as ``hyperfine ghosts''~\cite{bellos11,bellos12}. Since this population is spectroscopically distinct, we know that we are selecting only for $F = 3$ atoms in our initial photoassociation step. As we start with an initial population consisting of only $\left| F = 3 \right> + \left| F = 3 \right>$ molecules and end with a qualitatively equal mix of all three asymptotes after a two-photon process, up to one nuclear spin change must accompany each electronic transition.

So far in this analysis, we have discussed only the asymptotic spacing of the molecular hyperfine levels. Since we are actually dealing with bound states, it follows that the molecular hyperfine potentials must be essentially parallel to each other as they come in to shorter distances. Indeed, this is reflected in the new potential curves and vibrational eigenstates calculated by Tiemann~\cite{tiemann15}.

\begin{figure}[tbh]
\begin{center}
\includegraphics[width=\columnwidth]{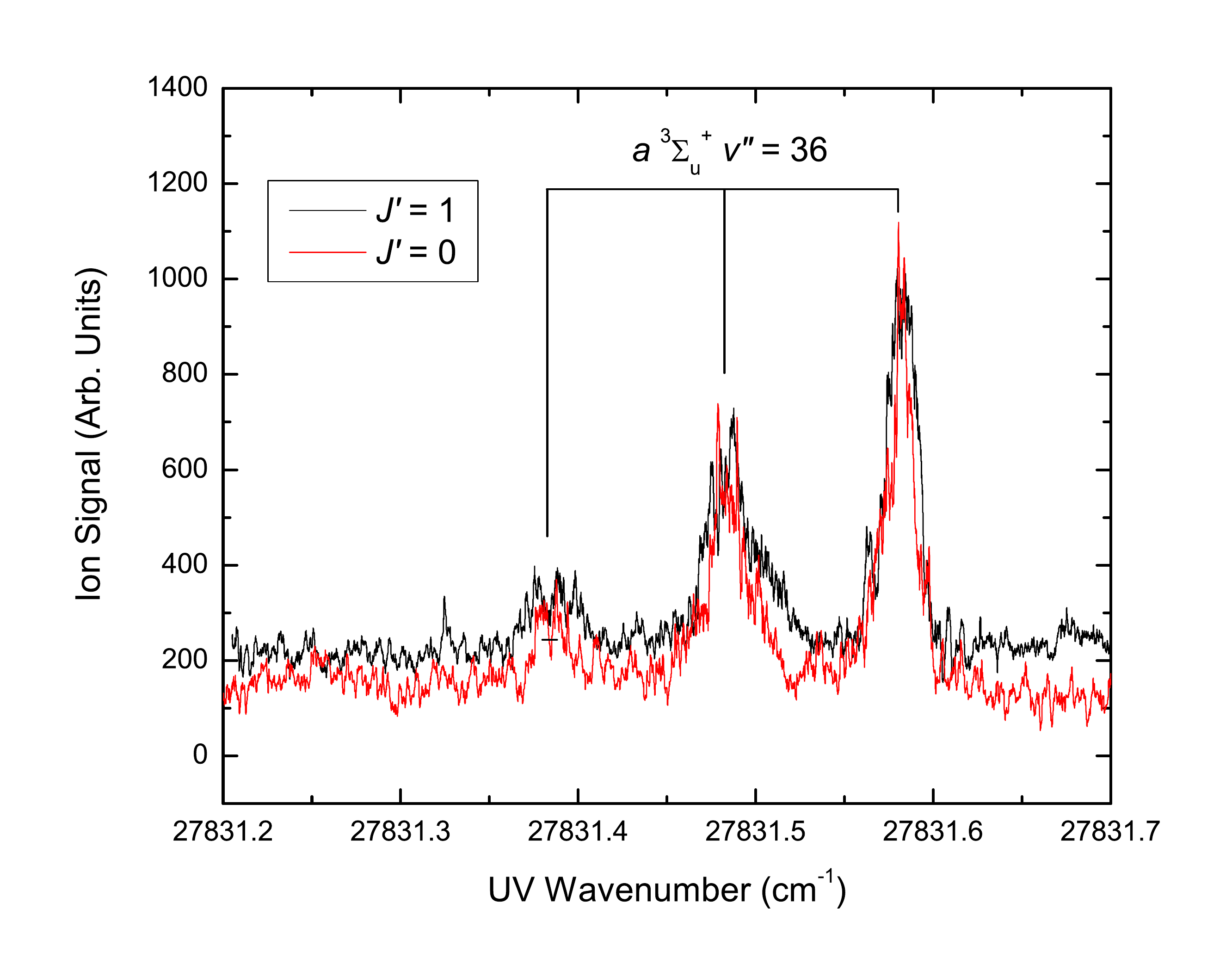}
\caption[Two scans of one inner-well line, with one scan using PA to a $J' = 1$ line and the other to a $J' = 0$ line.]{[Color online] Two scans of the same inner-well line are shown. In the first scan (black), the precursor $a \, ^3 \Sigma_u^+$ state molecules were formed by photoassociation to the $1 \, (0_g^-) \, v' = 173, \, J' = 1$ level. The second scan (red) was identical except that the photoassociation was locked to the $J' = 0$ level.}
\label{rydberg:jcomparison}
\end{center}
\end{figure}

Aside from the hyperfine distribution, we also must consider the rotational distribution. In general, we photoassociate to the $J' = 1$ level of a state because it has the greatest formation rate, and this is true of the $1 \, (0_g^-)$ $v' = 173$ level as well. From the selection rules given in Chapter 1.3 of~\cite{carollo15} we see that $\Delta J = \pm 1$ and $\Delta J = 0$ if $\Omega \neq 0$, which is true for the $\Omega = 1$ component of the $a \, ^3 \Sigma_u^+$ state. Thus for $J' = 1$, we can have $J'' = 0$, 1, or 2. If instead we photoassociate to the $J' = 0$ line, we can have $J'' = 0$ or 1. This comparison between $J' = 1$ and $J' = 0$ photoassociation was performed experimentally, and the result can be seen in Fig.~\ref{rydberg:jcomparison}.

\begin{figure}[tbh]
\begin{center}
\includegraphics[width=\columnwidth]{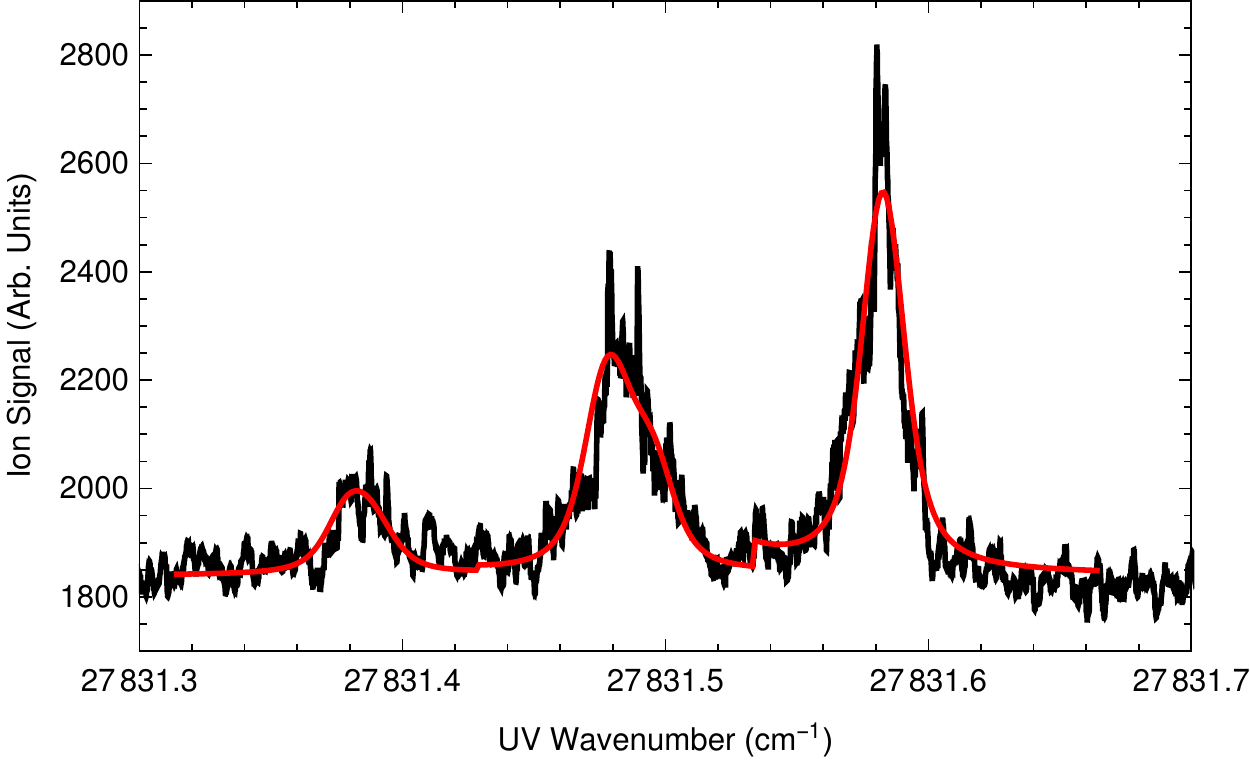}
\caption[The $J' = 0$ level from Fig.~\ref{rydberg:jcomparison} with a model fit]{[Color online] The $J' = 0$ level from Fig.~\ref{rydberg:jcomparison} is shown (black) with a model fit (red) based on semi-empirical data~\cite{tiemann15}. $N'' = 0$ and 1 are included. The model line heights are artificially scaled to match the data so that widths may more easily be compared.}
\label{rydberg:j0fit}
\end{center}
\end{figure}

\begin{figure}[tb]
\begin{center}
\includegraphics[width=\columnwidth]{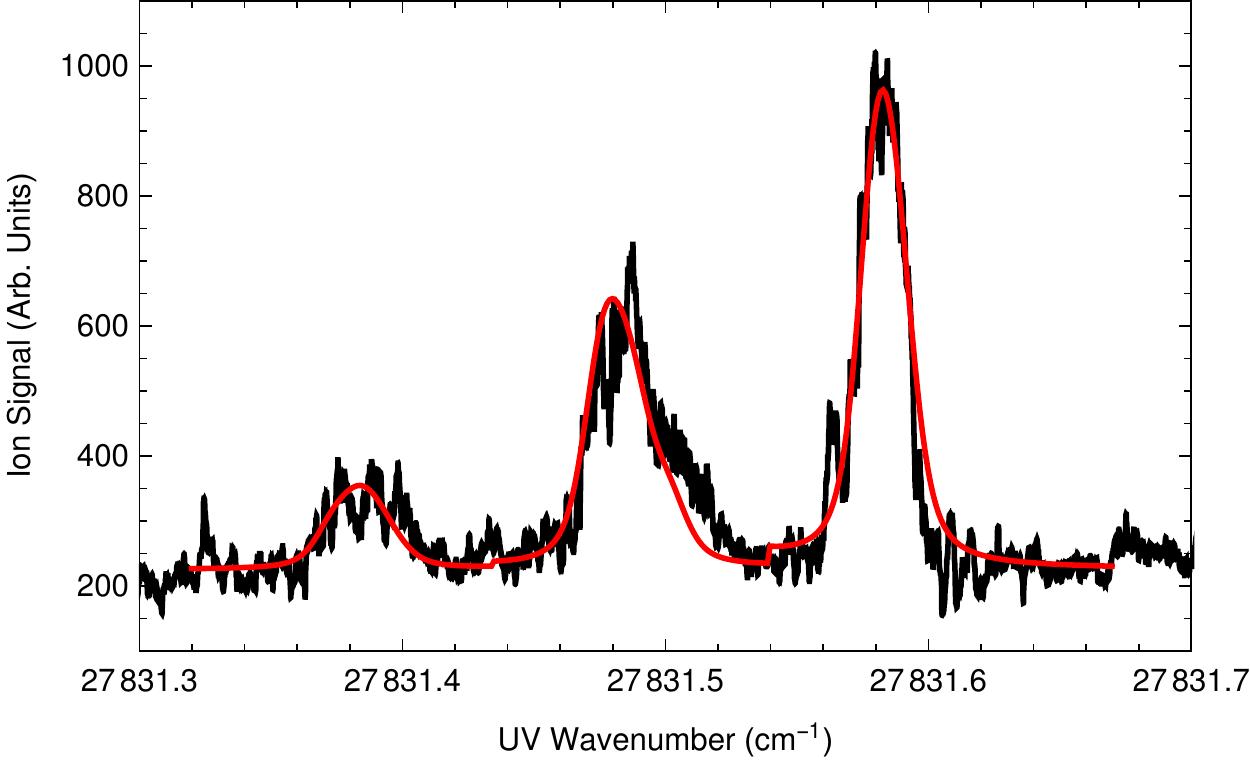}
\caption[The $J' = 1$ level from Fig.~\ref{rydberg:jcomparison} with a model fit]{[Color online] The $J' = 1$ level from Fig.~\ref{rydberg:jcomparison} is shown (black) with a similar model fit (red). Here $N'' = 0$, 1, and 2 are included, along with the same height scaling.}
\label{rydberg:j1fit}
\end{center}
\end{figure}

\begin{figure}[tbh]
\begin{center}
\includegraphics[width=\columnwidth]{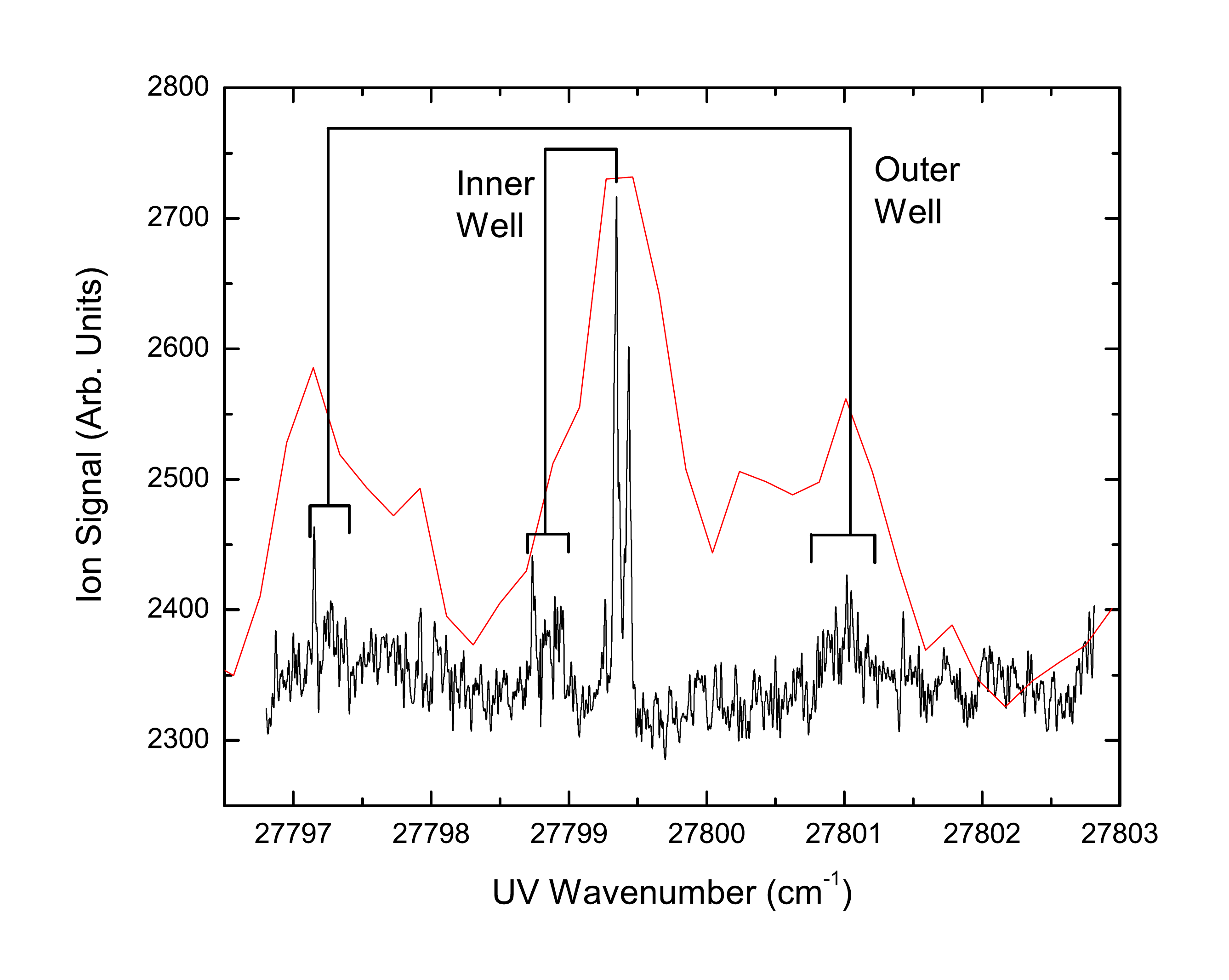}
\caption[A single spectrum showing an inner-well line flanked by two weaker outer-well levels.]{[Color online] A long scan was stitched together to show a single inner-well level with significantly-weaker outer-well levels to either side. The relative strength of these levels is qualitatively similar to the FCFs shown in Fig.~\ref{rydberg:overview}. The low-resolution data from Ref.~\cite{bellos13a} is seen in red. This data has been slightly shifted to match the current spectrum.}
\label{rydberg:long}
\end{center}
\end{figure}

The data from Tiemann~\cite{tiemann15} also includes rotational components, although in his case (e) basis set the appropriate quantum number is $N$, where $\mathbf{N} = \mathbf{J} - \mathbf{S} = \mathbf{L} + \mathbf{\ell}$. Rotational quanta $N'' = 0$, 1, and 2 were included to match the case of $J' = 1$ photoassociation. In $J' = 0$ photoassociation (including $N'' = 0$ and 1), we can see a narrower distribution in both the data and the model, which is shown in Figs.~\ref{rydberg:j0fit} and~\ref{rydberg:j1fit}. The Hamiltonian and computational method he uses for these calculations is described in Ref.~\cite{tiemann15a}, and his potentials and hyperfine functions are published in Ref.~\cite{tiemann10}. Our model assumes that the $a ^3 \Sigma_u^+$ state levels have a negligible linewidth due to their metastable nature. The line shapes are modeled by a Voigt profile, with the laser linewidth assumed to be Gaussian and the natural linewidth of the transitions taken to be Lorentzian.

Three scans of the inner well are modeled. From the models, we extract an average laser linewidth of 95 MHz. The modeled excited-state linewidth is an average of $\sim150$ MHz. Outer-well lines are more difficult to fit, due to lower signal-to-noise ratios, and an experimental artifact that appears to slightly shift the positions of some lines relative to each other. One of the outer-well lines fits well, giving a laser linewidth of 96 MHz, consistent with inner-well measurements, and an excited-state linewidth of 25(7) MHz (95\% confidence interval).

As some substructure must be present due to multiple hyperfine and rotational levels, this represents an overestimate of the linewidth. The molecular hyperfine is expected to follow the atomic hyperfine splitting, much as we see in the ground state. The hyperfine $A$ constant is given in Ref.~\cite{sansonetti06} for the $^{85}$Rb $7p_{1/2}$ state as $0.000590(3)$ cm$^{-1}$. For our values of $J = 1/2$, $I = 5/2$, and $F = 2$, 3 we find an atomic splitting of $\Delta E_{hf}/h = 3A = 0.00177$ cm$^{-1}$, or 53.1 MHz. With this splitting, the three molecular hyperfine asymptotes will extend over $\sim 106$ MHz, and the shorter-range potentials will vary from this in an unknown manner. Since we cannot fully account for the effect on the short-range potentials, we will continue to use 150 MHz as an upper bound on the excited-state linewidth.

Some fraction of the linewidth could also be due to radiative decay that competes with the autoionization, but we have no estimates of this contribution. Natural linewidth is related to lifetime by $\tau = \frac{1}{\omega}$ in rotational frequency units, which becomes $\tau = \frac{1}{2 \pi f}$ in Hz. A 150 MHz linewidth corresponds to an autoionization lifetime of $1.1 \times 10^{-9}$ s. As this is an upper bound to the linewidth, the autoionization lifetime must be a lower bound - the states could live longer. For comparison, the atomic $7p_{1/2}$ and $7p_{3/2}$ lifetimes are 272(15) ns and 246(10) ns, respectively~\cite{marek80}.

An interesting region of the spectrum is marked by the lowest green arrow in Fig.~\ref{rydberg:overview}. This  region, expanded in Fig.~\ref{rydberg:long}, contains both inner-well and outer-well states (marked by red arrows to either side in the figure), encouraging a direct comparison. The inner well, as before, has both $v'' = 35$ and $v'' = 36$ components, but now there is a strong asymmetry between them. The $v'' = 35$ component remains strong, while the $v'' = 36$ component is quite weak. We believe this is due to different Franck-Condon overlaps with the precursor molecular wavefunctions. As we explore more deeply-bound inner-well states, we are also probing the shorter internuclear distances that favor excitation of $v'' = 35$.

To either side of the inner-well levels in Fig.~\ref{rydberg:long}, we see weak features. As can be seen in the figure, these features correspond to lines seen in our low-resolution spectra (in red). Although we cannot easily resolve the structure of these features, they are too closely-spaced to an inner-well level to be from the inner well themselves. The characteristic spacing of the inner-well energy levels in this region is $\sim 5$ cm$^{-1}$, while the spacing of the outer-well levels is $\sim 2$ cm$^{-1}$. As each of these features is $\sim 2$ cm$^{-1}$ away from the inner-well line, they can only be outer-well states. The weak nature of these lines encourages us to look at much lower energies, at the bottom of the outer well, where the Franck-Condon factors are higher by roughly a factor of 40.

\begin{figure}[tbh]
\begin{center}
\includegraphics[width=\columnwidth]{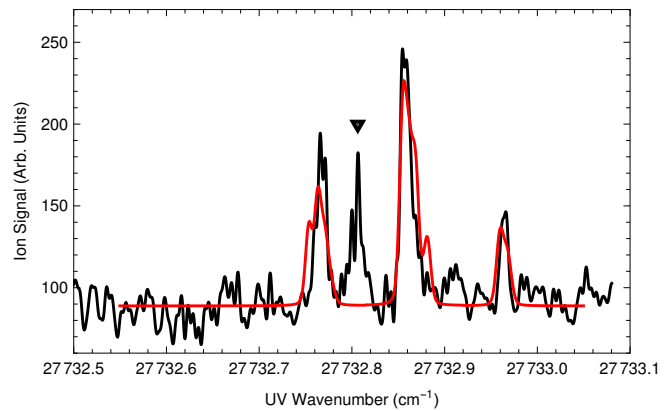}
\caption[$v' = 0$ of the outer well]{[Color online] A scan of the lowest-lying outer-well level (black) we have detected. We believe it is $v' = 0$ of the outer well. The overlay is a fitted model (red) with a laser linewidth of 96 MHz and a natural linewidth of 24.5 MHz.}
\label{rydberg:outer1}
\end{center}
\end{figure}

\begin{figure}[tbh]
\begin{center}
\includegraphics[width=\columnwidth]{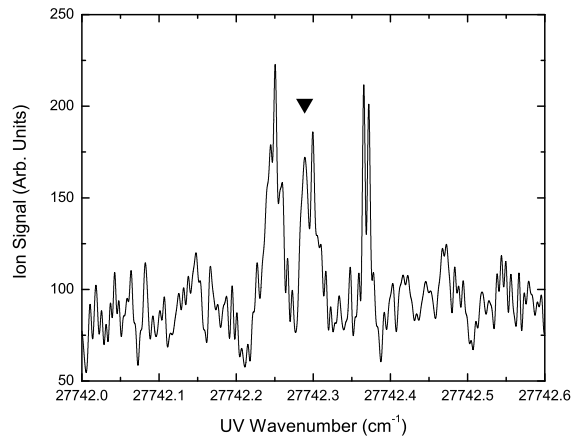}
\caption[$v' = 2$ of the outer well]{A scan of a slightly higher level of the outer well. We believe it is $v' = 2$ based on predicted line spacings.}
\label{rydberg:outer2}
\end{center}
\end{figure}

In our scans of the low-lying outer-well states, Figs.~\ref{rydberg:outer1} and~\ref{rydberg:outer2}, there is a spectral feature that does not match the ground-state hyperfine structure (indicated by a $\blacktriangledown$). This state is in nearly the same place in multiple spectra, $\sim 0.04$ cm$^{-1}$ higher in energy than the lowest-energy line.

At long range, the $X \, ^1 \Sigma_g^+$ state and the $a \, ^3 \Sigma_u^+$ state have the same van der Waals $C_6$, $C_8$, and $C_{10}$ coefficients~\cite{jones06}. This gives them a very similar vibrational progression at high $v''$ (where the outer turning point is at long range), and leads to a significant degree of degeneracy between levels of opposite symmetry. Despite weak interaction between these states, with such small energy detuning the states are almost completely coupled, as described in\ignore{Chapter~\ref{resonantcoupling}, e.g. in} Refs.~\cite{heinzen97,tiemann10}.

A calculation~\cite{leroylevel16} of energy levels using the potentials of Ref.~\cite{tiemann10} shows that the $X \, ^1 \Sigma_g^+$ state has levels $v'' = 118$ at $- 0.8351$ cm$^{-1}$  and $v'' = 119$ at $-0.4294$ cm$^{-1}$. These are, respectively, 0.0286 and 0.0133 cm$^{-1}$ more deeply bound than the $a \, ^3 \Sigma_u^+$ state $v'' = 35$ and 36 levels. These spacings, however, are in the completely decoupled basis set of non-interacting potentials. In the actual potentials, the significant interactions between the triplet and singlet states are represented in the model we used to create the fits in Figs.~\ref{rydberg:j0fit} and~\ref{rydberg:j1fit}.

However, these model fits do not show any extra line in the position observed. The $v'' = 34$ of the model does have a singlet-character line intermediate between two of the main groupings, but it is in between the wrong lines to match our data. Also, Franck-Condon factor calculations for $v'' = 34$ do not support strong transitions to the outer well. This is expected, due to the shorter internuclear separation of the outer lobe of the wavefunction.

One possible explanation for this extra line is that it comes from another potential curve in the same region. The most likely candidate for this is the $^3\Sigma_u^+$ state, the potential for which was calculated at the same time as the $^3\Sigma_g^+$ and is given in Ref.~\cite{bellos13}. Although $u \leftrightarrow u$ transitions are forbidden, the vibrational levels of the $a \, ^3\Sigma_u^+$ state that we use show strong singlet-triplet, and $g$-$u$, mixing, making such transitions possible.

\section{Conclusion}
We have acquired high-resolution spectra of the long-range Rydberg molecular state of $^{85}$Rb$_2$ near the $7p$ asymptote. This state has a short-range inner well and a long-range outer well, both of which have been observed. Much of the observed structure is due to coupling of states near the atomic ground-state asymptote, the hyperfine structure of which greatly complicates the spectrum. One spectral feature of the outer well continues to elude a definite assignment, on which investigations will continue. After modeling the remainder of the spectrum, we have placed a lower bound on the autoionization lifetime of the Rydberg molecule of $1.1 \times 10^{-9}$ s.

\acknowledgments
This research was funded with support from the National Science Foundation grant numbers PHY-1208317 and PHY-1506244 and Air Force Office of Scientific Research grant number FA9550-09-1-0588. We are particularly grateful to Prof. Eberhard Tiemann for his private communication concerning the hyperfine structure of the $X \, ^1\Sigma_g^+$ and $a \, ^3\Sigma_u^+$ states of $^{85}$Rb$_2$ near dissociation.

\bibliography{ultracold_references}

\end{document}